\documentclass[fleqn,twoside]{article}
\usepackage[centertags]{amsmath}
\usepackage{espcrc2}
\usepackage{graphicx}
\usepackage[figuresright]{rotating}

\title{Monopole Condensation in full QCD   using the
Schr\"odinger Functional}

\author{P. Cea\address{Dept. of Physics  and INFN - Bari - Italy},
        L. Cosmai\address{INFN - Bari - Italy}, and
        M. D'Elia\address{Dept. of Physics and INFN - Genova - Italy}}

\begin{document}

\begin{abstract}
We use a lattice thermal partition functional to study Abelian
monopole condensation in full QCD with $N_f=2$ staggered fermions.
We present preliminary results  on $16^3\times4$ and $32^3\times4$
lattices. \vspace{1pc}
\end{abstract}

\maketitle

\section{INTRODUCTION}

To give a possible explanation of color confinement
G.~'t~Hooft~\cite{tHooft:1976eps} and
S.~Mandelstam~\cite{Mandelstam:1974pi} suggested long time ago
that vacuum in gauge theories behaves like a magnetic (dual)
superconductor. The dual superconductivity hypothesis relies upon
the very general assumption that the dual superconductivity of the
ground state is realized if there is condensation of Abelian
magnetic monopoles.

\subsection{The thermal partition functional}

To investigate  vacuum structure  of lattice gauge theories at
finite temperature we introduced~\cite{Cea:2001an,Cea:2000zr} a
lattice thermal partition functional in presence of a static
background field
\begin{equation}
\label{ZetaTnew}  \mathcal{Z}_T \left[ \vec{A}^{\text{ext}}
\right] =
\int_{U_k(L_t,\vec{x})=U_k(0,\vec{x})=U^{\text{ext}}_k(\vec{x})}
\mathcal{D}U \, e^{-S_W}   \,,
\end{equation}
$S_W$ is the Wilson action, the physical temperature is given by
$T=1/a L_t$. The functional integration is extended over links on
a lattice $L_s^3 \times L_t$ with  hypertorus geometry  and
satisfying the constraints ($x_t$: temporal coordinate)
\begin{equation}
\label{coldwall}  U_k(x)|_{x_t=0} = U^{\mathrm{ext}}_k(\vec{x})
\,,\,\,\,\,\, (k=1,2,3) \,\,,
\end{equation}
$U^{\mathrm{ext}}_k(x)$ being the lattice version of the external
continuum gauge field $\vec{A}^{\mathrm{ext}}(x)=
\vec{A}^{\mathrm{ext}}_a(x) \lambda_a/2$.
\begin{equation}
\label{freeenergy} F[\vec{A}^{\mathrm{ext}}]= - \frac{1}{L_t} \ln
\frac{\mathcal{Z}_T[\vec{A}^{\mathrm{ext}}]} {{\mathcal{Z}}_T[0]}
\, .
\end{equation}
To circumvent the problem of computing a partition function which
is the exponential of an extensive quantity, in practical
computations we consider $F^{\prime}[\vec{A}^{\mathrm{ext}}]$ the
$\beta$-derivative of $F[\vec{A}^{\mathrm{ext}}]$. Indeed this
quantity  is simply related to  the plaquette $U_{\mu\nu}$
($\Omega = L_s^3 \times L_t$):
\begin{equation}
\label{deriv}
\begin{split}
F^{\prime}[\vec{A}^{\mathrm{ext}}]   & = \left \langle
\frac{1}{\Omega} \sum_{x,\mu < \nu}
\frac{1}{3} \,  \text{Re}\, {\text{Tr}}\, U_{\mu\nu}(x) \right\rangle_0  \\
& - \left\langle \frac{1}{\Omega} \sum_{x,\mu< \nu} \frac{1}{3} \,
\text{Re} \, {\text{Tr}} \, U_{\mu\nu}(x)
\right\rangle_{\vec{A}^{\mathrm{ext}}} \,,
\end{split}
\end{equation}
where the subscripts on the averages indicate the value of the
external field.

\subsection{Including fermions}

We are interested in gauge systems at thermal equilibrium in
presence of a static (time-independent) external background field.
So that  in presence of dynamical fermions   Eq.~(\ref{ZetaTnew})
becomes:
\begin{equation}
\label{ZetaTfermions}
\begin{split}
& \mathcal{Z}_T \left[ \vec{A}^{\text{ext}} \right]  = \\
& \int_{U_k(L_t,\vec{x})=U_k(0,\vec{x})=U^{\text{ext}}_k(\vec{x})}
\mathcal{D}U \,  {\mathcal{D}} \psi  \, {\mathcal{D}} \bar{\psi}
e^{-(S_W+S_F)}   \,,
\end{split}
\end{equation}
where we integrate on fermionic fields  without any constraint. As
usual we impose on fermionic fields
 periodic boundary conditions in the spatial directions and
antiperiodic boundary conditions in the temporal direction.

\subsection{Detecting monopole condensation}

Abelian monopole condensation can be detected using order/disorder
parameters~\cite{Kadanoff:1971kz,Fradkin:1978th,DiGiacomo:1997sm}. The disorder
parameter is related to the monopole free energy and is defined by
means of the thermal partition function in presence of the Abelian
monopole background field:
\begin{equation}
\label{disorderT}  \mu = e^{-F_{\text{mon}}/T_{\text{phys}}} =
\frac{\mathcal{Z}_T \left[ \mathbf{A}^{\text{mon}} \right]}
{\mathcal{Z}_T[0]} \,,
\end{equation}
$F_{\text{mon}}$ is the free energy to create a monopole. If there
is condensation  $F_{\text{mon}}$ is finite and $\mu \ne 0$. Note
that since our disorder parameter $\mu$ has been defined in terms
of the thermal partition functional $\mathcal{Z}_T $ that is
gauge-invariant for time-independent gauge transformations of the
external background fields {\em{gauge fixing is not needed}} to
perform the Abelian projection in the case of Abelian background
fields.

\section{ABELIAN MONOPOLES IN FULL QCD}

\subsection{Definition}

In the continuum the magnetic monopole field with the Dirac string
in the direction  $\vec{n}$  is
\begin{equation}
\label{monopu1}
  e \vec{b}({\vec{x}}) =  \frac{n_{\mathrm{mon}}}{2}
\frac{ \vec{x} \times \vec{n}}{|\vec{x}|(|\vec{x}| -
\vec{x}\cdot\vec{n})} \,,
\end{equation}
where, according to the Dirac quantization condition,
 $n_{\mathrm{mon}}$ is an integer and
$e$  is the electric charge  magnetic charge =
$n_{\mathrm{mon}}/2e$.

For SU(3) gauge theory the maximal Abelian group is
U(1)$\times$U(1), therefore we may introduce two independent types
of Abelian monopoles associated respectively to the
 $\lambda_3$  and  $\lambda_8$
diagonal generators of SU(3). We shall consider here  the
$\lambda_3$  Abelian monopole field. On the lattice it is given by
($(X_1,X_2,X_3)$ monopole coordinates)
\begin{eqnarray}
\label{t3linkssu3}
%
U_{1,2}^{\text{ext}}(\vec{x}) & =
\begin{bmatrix}
e^{i \theta^{\text{mon}}_{1,2}(\vec{x})} & 0 & 0 \\ 0 &  e^{- i
\theta^{\text{mon}}_{1,2}(\vec{x})} & 0 \\ 0 & 0 & 1
\end{bmatrix}
\,, \nonumber \\ U^{\text{ext}}_{3}(\vec{x}) & = {\mathbf 1} \,, \\
\theta^{\text{mon}}_{1,2}(\vec{x}) & = \mp
\frac{n_{\text{mon}}}{4}
\frac{(x_{2,1}-X_{2,1})}{|\vec{x}_{\text{mon}}|}
\frac{1}{|\vec{x}_{\text{mon}}| - (x_3-X_3)} \,. \nonumber
\end{eqnarray}

\subsection{Numerical simulations}

We used the standard HMC R-algorithm for two degenerate flavors of
staggered fermions with a quark mass $a m = 0.075$ (at this value
of the mass and $N_T=4$ $\rightarrow$ $\beta_c \sim
5.35$~\cite{Aoki:1998wg}). We have collected about 2000
thermalized trajectories for each value of $\beta$ at $L_s =16$
and about 500 thermalized trajectories for each value of $\beta$
at $L_s=32$. Each trajectory consists of $125$  molecular dynamics
steps and has total length  $1$. The computer simulations have
been performed on the APEmille crate at INFN/Bari.

\subsection{Numerical results}
\begin{figure}[t]
\begin{center}
\includegraphics[width=0.5\textwidth,clip]{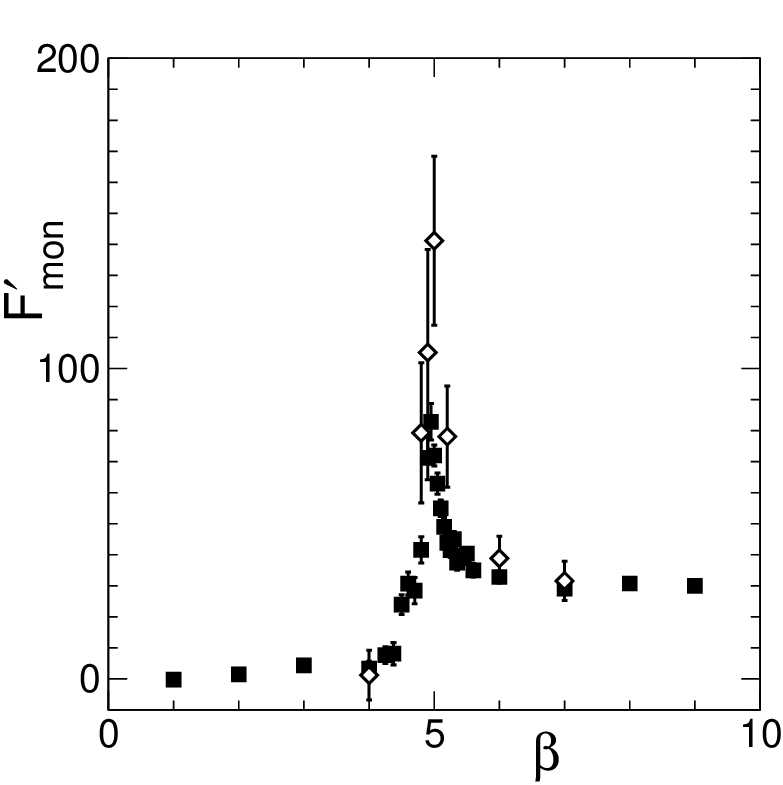}
\vspace{-1.1cm} 
\caption{The $\beta$-derivative of the monopole
free energy $F^{\prime}_{\text{mon}}$ on a $L_s^3 \times 4$ lattice
for 2 flavors QCD (full squares and open diamonds refer
respectively to $L_s=16$ and $L_s=32$).}
\end{center}
\end{figure}
\begin{figure}[t]
\begin{center}
\includegraphics[width=0.5\textwidth,clip]{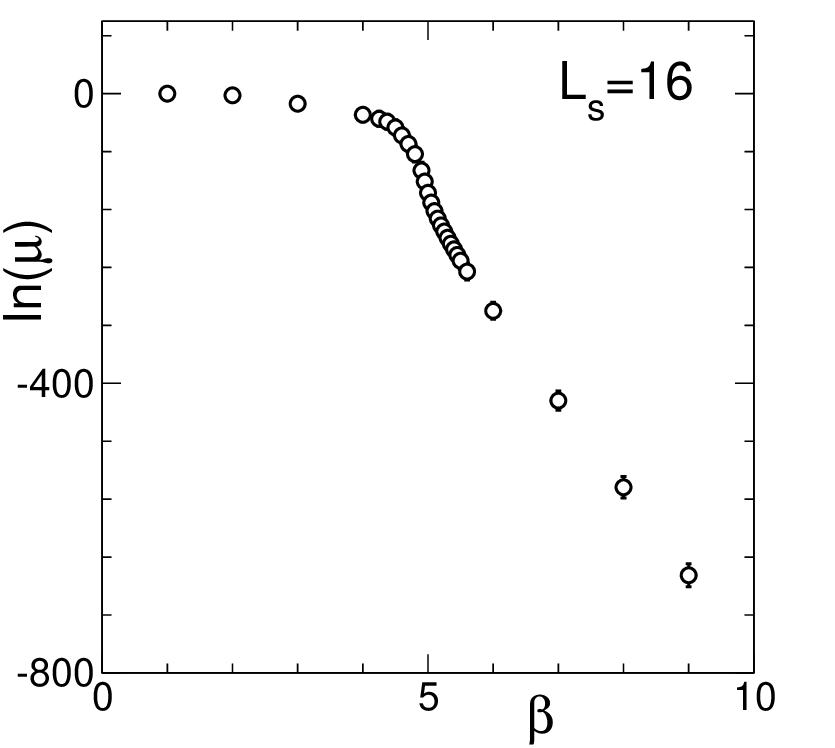}
\vspace{-1.1cm} \caption{The logarithm of the disorder parameter
Eq.~(\ref{disorderT}) for 2 flavors QCD and $L_s=16$.}
\end{center}
\end{figure}

In Fig.~1 we compare $F^{\prime}_{\text{mon}}$ in the 2 flavors full
QCD case for two different spatial volumes. Data in the weak and
strong coupling regions agree within statistical errors. On the
other hand the signal at the peak value gets increased going to a
larger spatial volume.

Observing that $F_{\text{mon}}= 0$  at  $ \beta = 0$, we may
obtain $F_{\text{mon}}$  from $F^{\prime}_{\text{mon}}$  by a
numerical integration in $\beta$ and from Eq.~(\ref{disorderT}) we
are able to estimate the disorder parameter $\mu$. Fig.~3 shows
that  $\ln \mu = 0$ in the confined phase, i.e. the free energy
required to create an Abelian monopole in the QCD vacuum is zero
and therefore Abelian monopoles condense in the confined phase.
\begin{figure}[t]
\begin{center}
\includegraphics[width=0.5\textwidth,clip]{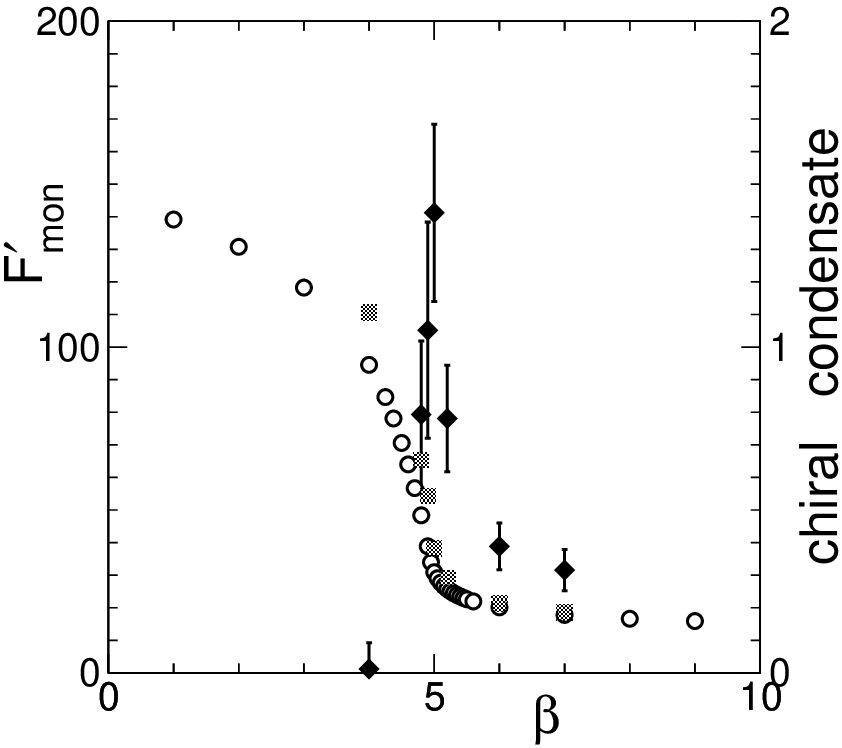}
\vspace{-1.1cm} \caption{$F^{\prime}_{\text{mon}}$  in 2 flavors QCD
and $L_s=32$ (full diamonds) compared with chiral condensate at
$L_s=16$ (open circles) and $L_s=32$ (full grey squares).}
\end{center}
\end{figure}

Fig.~4 displays  $F^{\prime}_{\text{mon}}$  in the 2 flavors full
QCD case and $L_s=32$  compared with the chiral condensate. The
peak in  $F^{\prime}_{\text{mon}}$  corresponds to the drop of the
chiral condensate.

\section{CONCLUSIONS}

Using a  thermal partition functional in presence of an external
Abelian monopole background field we simulate 2 staggered flavors
full QCD on a $L_s^3 \times 4$ lattice. We find that Abelian
monopoles condense in the spontaneously broken chiral symmetry
phase of  2 flavors full QCD. This
is consistent with similar results found in Ref.~\cite{Carmona:2002ty}
For a better understanding of
Abelian monopole condensation in full QCD we expect to perform new
simulations at different  values of the masses and of the flavor
numbers as well a finite size scaling analysis.


\end{document}